\documentclass[11pt]{article}
\usepackage{graphicx}
\usepackage{sao1}
\newcommand{\eps}[1]{\log\varepsilon_{\rm #1}}
\newcommand{\kms}{km\,s$^{-1}$}
\newcommand{\Vmic}{$V_{\rm mic}$}
\newcommand{\Teff}{T_{\rm eff}}
\newcommand{\Eexc}{$E_{\rm exc}$}
\newcommand{\logg}{\rm log~ g}

\begin{document}

\title{Non-LTE line formation for Fe~I and Fe~II \\ in the atmospheres of A-F type stars}

\author{L. Mashonkina\inst{ } }
\institute{Institute of Astronomy, Russian Academy of Sciences, RU-119017 Moscow, Russia
}
\maketitle 

\begin{abstract}
Non-local thermodynamical equilibrium (non-LTE) line formation for neutral and singly-ionized iron is considered through a range of spectral types when the effective temperature varies from 6500~K up to  8500~K, the gravity from $\logg = 4$ down to $\logg = 3$, and the metallicity is solar one. The non-LTE calculations were performed with a comprehensive model atom for iron which was treated in our earlier paper (Mashonkina~et~al. 2011). The departures from LTE lead to systematically depleted total absorption in the Fe~I lines and positive abundance corrections, in qualitative agreement with the Rentzsch-Holm (1996) study. However, our predicted magnitude of the non-LTE effects is significantly smaller compared to the previous results due to the use of rather complete model atom of Fe~I. The non-LTE abundance corrections do not exceed 0.1\,dex for the dwarf models and 0.20~dex for the giant ones. Non-LTE leads to strengthening the Fe~II lines, however, the effect is small, such that the abundance correction is at the level of $-0.01$ to $-0.03$\,dex over the whole range of stellar parameters being considered. No firm conclusion can be drawn with respect to whether or not the Fe~I/Fe~II ionization equilibrium is fulfilled in the atmosphere of the Sun and Procyon due to uncertainty in available $gf-$values for visible lines of Fe~I and Fe~II. 
\keywords{Atomic data -- Line: formation -- Stars: atmosphere}
\end{abstract}

\section{Introduction}

 Iron is the most popular chemical element in studies of both non-magnetic and magnetic A and later type stars thanks to quite numerous lines of the neutral and singly-ionized species in the visible spectrum. Iron lines are used to determine stellar parameters, such as effective temperature, $\Teff$, surface gravity, $\logg$, microturbulence velocity, \Vmic, and, in the magnetic stars, also to investigate magnetic fields. Therefore, the calculations of the iron lines have to be based on realistic line formation scenarios. 

   In the atmospheres with $\Teff >$ 4500~K, Fe~I is a minority species. Number density of the minority species can easily deviate from the thermodynamical equilibrium (TE) population due to small deviation of the mean intensity of ionizing radiation from the Planck function. Since the beginning of 1970th a number of studies attacked a problem of the non-local thermodynamic equilibrium (non-LTE) line formation for Fe~I in the Sun and late type stars (Tanaka, 1971; Athay \& Lites, 1972; Boyarchuk et al., 1985; Takeda, 1991; Gratton et al., 1999; Th\'evenin \& Idiart, 1999; Gehren et al., 2001; Shchukina \& Trujillo Bueno, 2001; Collet et al., 2005) Here we list only the studies, where original model atom was produced. The non-LTE effects in the hotter atmospheres were investigated by only Gigas (1986) and Rentzsch-Holm (1996). 
All these research came to a common conclusion that Fe~I is subject to overionization in stellar atmospheres due to enhanced photoionization of the levels with the ionization edges in the ultra-violet (UV) and the Fe~I lines are weakened compared to their LTE strengths. However, a consensus on the expected magnitude of the non-LTE effects was not reached. For example, the surface gravity correction for cool metal-poor atmospheres was calculated, in some studies, at the level of 0.5~dex, while no significant departures from LTE were found by others. The discrepancies between different papers might be due to an incompleteness of the applied model atoms. 

Mashonkina et al. (2011) constructed a comprehensive model atom for iron using more than 3000 measured and predicted energy levels and showed that the departures from LTE for Fe~I decreased significantly in the atmospheres of solar-type stars, in particular, in the metal-poor atmospheres compared to those calculated in the previous non-LTE studies.
In this study, we revise the non-LTE effects for the stellar parameter range characteristic of A-type stars. The paper is organized as follows. An introduction to the non-LTE approach is given in the next section. Sect.\,\ref{sect:NLTE} describes briefly the method of the non-LTE calculations for Fe~I-Fe~II. The results for the two reference stars, the Sun and Procyon, are presented in Sect.\,\ref{sect:sun}. Finally, we report on the departures from LTE depending on stellar parameters.

\section{What is meant by a non-LTE approach?}

 We assume that the particle velocity distribution is Maxwellian with a single kinetic temperature for various kind particles, i.e., for electrons, atoms, and ions: $T_e = T_a = T_i$. We consider that the occupation number of any atomic level is determined from the balance between radiative and collisional population and de-population processes, such as  
photoexcitation, photoionization and their inverse processes, inelastic collisions with  electrons, atoms, molecules, dielectronic recombination, charge exchange, i.e., from a
statistical equilibrium (SE): 

\begin{equation}
n_i \sum_{j \ne i}(R_{ij} + C_{ij}) = \sum_{j \ne i}n_j (R_{ji} +
C_{ji}) ~~~~~~~~~~i=1, \ldots , NL. \label{SE}
\end{equation}

\noindent Collision rate  $C_{ij}$ is defined by the local kinetic temperature $T_e$ and collider particle number density. Collision processes tend to establish TE. Radiative rate $R_{ij}$ depends on the radiation
field, which is, in general, highly non-local in character. Radiative processes tend to destroy TE. 

In the nature, each chemical species possesses a huge number of bound energy levels approaching to the infinity. In practice, the real atomic term structure is represented by the model atom with finite number of levels $NL$. To investigate the SE of an atom, a large amount of various atomic data is required, such as energy levels, photoionization cross-sections for every level in the model atom, oscillator strengths ($f_{ij}$) for the entire set of allowed transitions, and collision excitation and ionization data for the entire set of transitions. 
For the last two decades, the situation with atomic data has improved
significantly. A fairly extensive set of accurate atomic data on
photoionization cross-sections and oscillator strengths was calculated in the Opacity Project (Seaton~et~al., 1994) and has become available via the TOPBASE database. The IRON project is in progress and provides the data for the Fe group
elements. Quantum-mechanical {\it ab initio} calculations exist for electron impact excitation in a number of selected atoms and ions. However, for heavy elements beyond the Fe group, much data are missing, in particular, for the high-excitation states. Rough theoretical approximations are still used to evaluate inelastic collision cross-sections. The calculated SE of an atom depends on the completeness of the adopted model atom and the accuracy of the used atomic data. 

The excitation and ionization states of the matter is calculated from the solution of combined SE (\ref{SE}) and radiation transfer equations (\ref{RT}): 

\begin{equation}
\mu\frac{d I_\nu(z,\mu)}{d z} = -\chi_\nu(z) I_\nu(z,\mu) + \eta_\nu(z)
\label{RT}
\end{equation}

\noindent Here, $\chi_\nu$ and $\eta_\nu$ are the absorption and emission coefficients, respectively. The radiation transfer equations have to be solved at frequences of all the radiative transitions in the atom which makes the non-LTE calculations to be bulky. 
In non-LTE, atomic level population $ n_i(d)$ at any depth point
$d$ depends on the physical conditions throughout the atmosphere: $ n_i(d) = f(n_1, \ldots, n_{NL}, J_1, \ldots,
J_{NF})$. Here, $J_{\nu}$ is the mean intensity.
The models in which the Saha-Boltzmann equations are replaced by the physically more accurate SE equations are called the non-LTE models.

 For comparison, in LTE, level populations are calculated from the Saha and Boltzmann equations and depend only on local temperature and electron number density. This implies simple numerical calculations and the need in atomic data only for the spectral line under investigation. 
     The LTE assumption is valid if every transition in the atom is in detailed balance. This is not fulfilled in spectral line formation layers, where the mean-free path of photons is large and the intensity of radiation is far from being in TE. To evaluate the effect of the departures from LTE on line strengths and stellar parameters derived from these lines, one needs to solve the non-LTE problem for a given chemical species. 

\section{The method of non-LTE calculations for iron}\label{sect:NLTE}

The non-LTE calculations were performed using a revised version of the DETAIL program (Butler \&  Giddings, 1985) based on the accelerated lambda iteration (ALI) scheme. 

\paragraph{Model atom of Fe~I-Fe~II.}  
We applied the model atom treated by Mashonkina et al. (2011). It was constructed using the 958 energy levels of Fe~I known the experimental analysis of Nave~et~al. (1994) and also the levels predicted from calculations of the Fe~I atomic structure by Kurucz (2009), in total 2970 levels. The system of measured levels is nearly
complete below \Eexc\, $\simeq 5.6$~eV. However, laboratory experiments do not see most of the high-excitation levels with \Eexc\ $>$ 7.1~eV, which should contribute a lot to provide close collisional coupling of Fe~I levels to the Fe~II ground state. Neglecting the multiplet fine structure gives 233 levels of Fe~I in the model atom. The predicted high-excitation levels were used to make up six super-levels. This is new compared to all previous non-LTE studies for Fe~I. In total, 11958 allowed transitions occur in this model atom of Fe~I. For Fe~II, we use the 89 lowest terms with \Eexc\ up to 10~eV. The ground state of Fe~III completes the system of levels in the model atom. 

Photoionization is the most important process deciding whether the Fe~I atom tends to depart from LTE in the atmosphere of a cool star. Our non-LTE calculations rely on the photoionization cross-sections of the IRON project (Bautista~et~al., 1997). The hydrogenic approximation was used for the Fe~II states. Radiative bound-bound (b-b) transition rates were computed using $gf$-values from the Nave~et~al. (1994) compilation and Kurucz (2009) calculations. All levels in our model atom are coupled via collisional excitation and ionization by electrons. The collisional rates were calculated using the theoretical approximation of van Regemorter (1962) and Seaton (1962). 

\paragraph{Model atmospheres.}  

For the Sun and Procyon, the calculations were performed with plane-parallel and blanketed LTE model atmospheres computed with the MAFAGS-OS code (Grupp et al., 2009), which is based on up-to-date continuous opacities and
includes the effects of line-blanketing by means of opacity sampling. Small grid of model atmospheres with $\Teff  = 6500-8500$~K and $\logg = 3$ and 4 was taken from the Kurucz (2008) website to evaluate the departures from LTE depending on stellar parameters.

\begin{figure}
\begin{center}
\hbox{
\hspace{-6mm}
\resizebox{170mm}{!}{\rotatebox{0}{\includegraphics{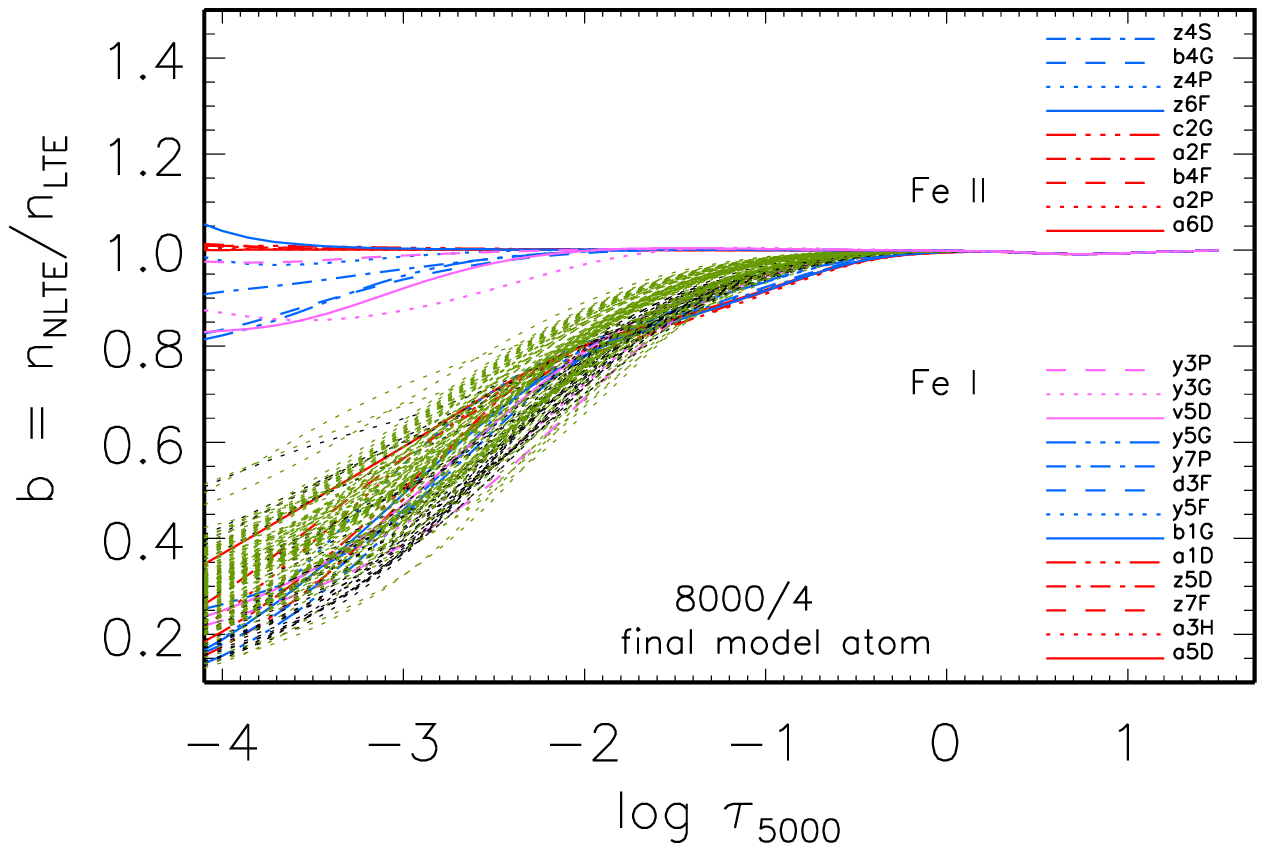}}
\hspace{-8mm} \rotatebox{0}{\includegraphics{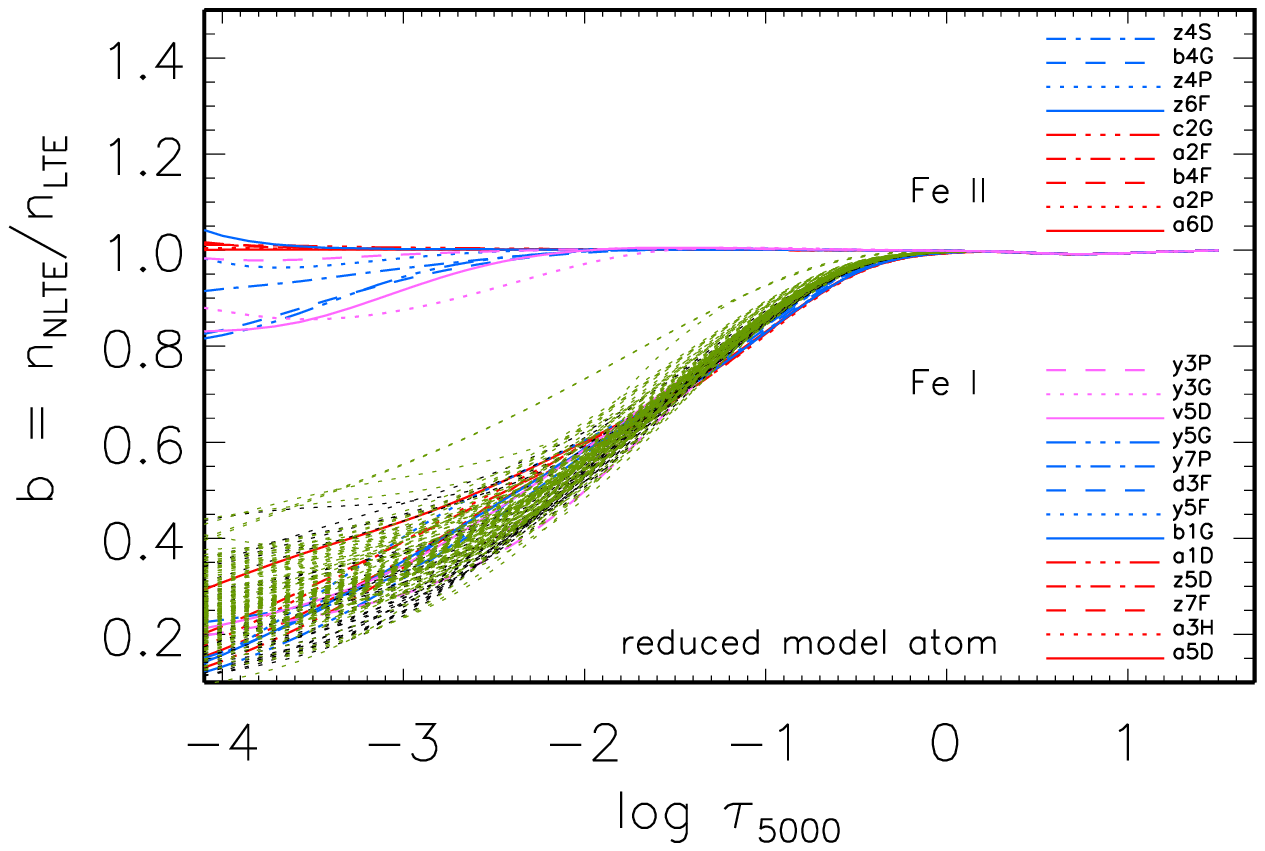}}}}
\caption{Departure coefficients, $b$, for the levels of Fe~I and Fe~II as a function of $\log \tau_{5000}$ in the model atmosphere 8000/4/0 from the calculations with our final model atom (left panel) and the reduced model atom which ignores the predicted levels of Fe~I (right panel).}
\label{bf_8000}
\end{center}
\end{figure}

\paragraph{Statistical equilibrium of iron.}  

The departure coefficients, $b_i = n_i^{\rm NLTE}/n_i^{\rm LTE}$, for the levels of Fe~I and Fe~II in the representative model atmosphere $\Teff$/$\logg$/[M/H] = 8000/4/0 are presented in Fig.\,\ref{bf_8000} as a function of $\log \tau_{5000}$. Here,
$n_i^{\rm NLTE}$ and $n_i^{\rm LTE}$ are the statistical
equilibrium and thermal (Saha-Boltzmann) number densities, respectively. The difference between two panels of Fig.\,\ref{bf_8000} is in using two different atomic models, i.e., the final model atom from Mashonkina~et~al. (2011) in the left panel and the reduced model atom which ignores the predicted levels of Fe~I in the right panel. Our calculations support qualitatively the previous results of Rentzsch-Holm (1996) for common stellar parameters.

\begin{itemize}
\item All the levels of Fe~I are underpopulated in the atmospheric layers above $\log \tau_{5000} = -0.3$ due to the overionization caused by superthermal radiation of a non-local origin below the thresholds of the low excitation levels of Fe~I.
\item Fe~II dominates the element number
density over all atmospheric depths. Thus, no process seems to affect the Fe~II ground-state and low-excitation-level populations significantly, and they keep their thermodynamic equilibrium values.
\end{itemize}

In this study, progress was made in establishing close collisional coupling
the Fe~I levels near the continuum to the ground state of Fe~II, due to including in the model atom a bulk of the predicted high-excitation levels of Fe~I. As a result, the populations of the Fe~I levels obtained with the final model atom (Fig.\,\ref{bf_8000}, left panel) are closer to the corresponding TE populations in the line-formation layers ($\log \tau_{5000} = 0$ up to $-3$) than that for the reduced model atom (Fig.\,\ref{bf_8000}, right panel). 

\section{Fe I/Fe II ionization equilibrium in the Sun and Procyon}\label{sect:sun}

The non-LTE method was applied to analyze the Fe~I and Fe~II lines in the two stars with well-determined stellar parameters: the Sun and Procyon. The solar flux observations are taken from the Kitt Peak Solar Atlas (Kurucz~et~al., 1984). The spectroscopic observations for Procyon were carried out by Korn (2003) with the fibre-fed \'echelle spectrograph FOCES at the 2.2m telescope of the Calar Alto Observatory, with a spectral resolving power of $R \simeq 60\,000$ and a signal-to-noise of $S/N \ge 200$. 

In the solar spectrum, we selected 54 unblended lines of Fe~I and 18 lines of Fe~II which cover a broad range of the line strength and excitation energy of the lower level. 
Despite the existence of many sources of $gf-$values for neutral iron, there is no single source that provides data for all the selected Fe~I lines. We employ experimental $gf-$values from Bard et al. (1991), Bard \& Kock (1994), Blackwell~et~al. (1979; 1982a; 1982b), Fuhr~et~al. (1988), O'Brian et al. (1991). Multiple sources of  $gf-$value are available for each line of Fe~II. We inspected 5 sets of data from Mel\'endez \& Barbuy (2009, hereafter MB09), Moity (1983, hereafter M83), Raassen \& Uylings (1998, hereafter RU98), Schnabel~et~al. (2004, hereafter SSK04), and the VALD database (Kupka~et~al., 1999). Van der Waals broadening of the iron lines is accounted for using the most accurate data available from calculations of Anstee \& O'Mara (1995), Barklem \& O'Mara (1997), Barklem et al. (1998), and Barklem \& Aspelund-Johansson (2005). 

\paragraph{The Sun.} 
With the solar model atmosphere 5780/4.44, the non-LTE effects are small for the Fe~I lines and negigible for the Fe~II lines, such that the non-LTE abundance correction amounts, on average, $\Delta_{\rm NLTE} = \eps{NLTE}-\eps{LTE}$ = +0.03~dex for Fe~I and smaller than 0.01~dex for Fe~II. A depth-independent microturbulence of 0.9\,\kms\, was adopted in the calculations. The iron abundance was derived from line profile fitting. For the Fe~I lines, we obtained the mean non-LTE abundance $\eps{FeI} = 7.56\pm0.09$. The statistical error which is defined as the dispersion in the single line measurements about the mean, $\sigma = \sqrt{\Sigma(\overline{x}-x_i)^2/(n-1)}$, is considerably too high. This is, most probably, due to the uncertaunty in $gf-$values because the abundance scatter is largely removed in a line-by-line differential analysis of solar type stars as shown by Mashonkina~et~al. (2011). For example, the Fe~I based abundance [Fe/H]$_I = 0.11\pm0.03$ for $\beta$~Vir (6060/4.11). 
 We find that the average Fe~II based abundance depends significantly on the used source of $gf-$values: $\eps{FeII} = 7.41\pm0.11$ (SSK04), $7.45\pm0.07$ (VALD), $7.47\pm0.05$ (MB09), and $7.56\pm0.05$ (RU98, M83). It is worth noting that the most recent laboratory $gf-$values of SSK04 lead to the highest abundance scatter. 

Large statistical errors and significant systematic discrepancies between different authors make the situation with oscillator strengths for the visible Fe~I and Fe~II lines to be unacceptable. Independent of either 1D or 3D modelling, no firm conclusion can be drawn with respect to whether or not the Fe~I/Fe~II ionization equilibrium is fulfilled in the solar atmosphere and how realistic the non-LTE calculations for iron are. The astrophysical community shall ask atomic spectroscopists for new accurate measurements.

With the cited $gf-$values for Fe~I and the data of RU98 for Fe~II, the ionization equilibrium between Fe~I and Fe~II is matched consistently with the solar $\logg = 4.44$. Is it possible to achieve a similar consistency for any other star using the same line list and the same atomic data?

\begin{figure}
\begin{center}
\includegraphics[width=120mm]{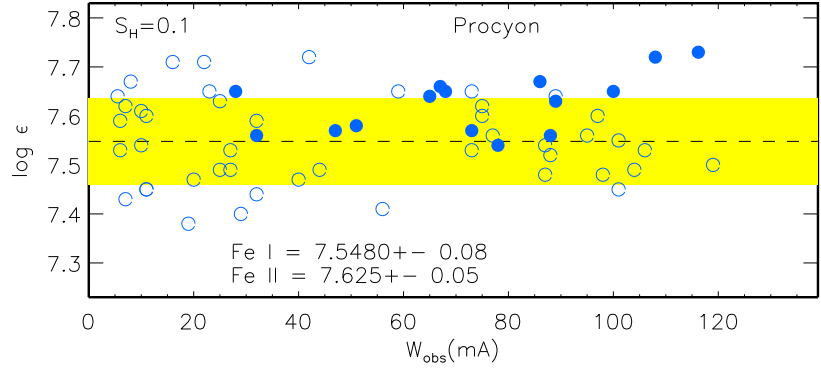}
\caption{Non-LTE iron abundances from the Fe~I (open circles) and Fe~II (filled circles) lines in Procyon plotted as a function of $W_{obs}$. The dashed line indicates the mean abundance derived from the Fe~I lines and the shaded grey area its statistical error. For Fe~II, $gf-$values are from Raassen and Uylings (1998).}
\label{fe_procyon}
\end{center}
\end{figure}

\paragraph{Procyon.}
 Procyon is among the very few stars for which the whole set of fundamental stellar parameters except metallicity can be determined from (nearly) model-independent methods. Allende Prieto~et~al. (2002) obtained $\Teff$ = 6510$\pm49$~K and $\logg$ = 3.96$\pm0.02$. Microturbulence velocity of \Vmic\,=1.6~\kms\ was determined in this study from the requirement that the non-LTE iron abundance derived from Fe~I lines must not depend on the line strength.

We find that the mean LTE abundance determined from the Fe~I lines is 0.14~dex lower than that from the Fe~II lines. The non-LTE abundances from the individual Fe~I and Fe~II lines are presented in Fig.\,\ref{fe_procyon}, with the mean values of $\eps{FeI} = 7.55\pm0.09$ and $\eps{FeII} = 7.62\pm0.06$. Non-LTE removes, in part, a disparity between two ions and leads to consistent within the error bars abundances from two ionization stages. But the abundance error is as large as for the Sun, and this leaves a space for speculations on either one needs to improve further the non-LTE line formation modelling, or to revise upwards $\Teff$ (change in $\Teff$ by 80~K fully removes a disparity between Fe~I and Fe~II), or to apply a new generation model atmosphere based on hydrodynamic calculations.

\begin{table}
\caption{Non-LTE abundance corrections (dex) for the selected lines of Fe~I and Fe~II. Multiplet numbers are indicated in the third string.}\label{tab_corr}
\begin{center}
\begin{tabular}{ccrrrrrcrrrr}\hline\noalign{\smallskip}
$\Teff$ & $\logg$ & \multicolumn{5}{c}{Fe I} & \ \ & \multicolumn{4}{c}{Fe II}\\
\cline{3-7}
\cline{9-12}
        &         & 5434 & 4920 & 5282 & 5217 & 5367 & & 4924 & 4583 & 5325 & 6247 \\
        &         & (15) & (318)& (383)& (553)& (1146)& & (42) & (37) & (49) & (74) \\
\hline\noalign{\smallskip}
   6500 & 4       & 0.06 & 0.08 & 0.12 & 0.12 & 0.04 & &-0.02 & 0.00 & 0.00 & -0.01 \\
   7000 & 4       & 0.06 & 0.06 & 0.10 & 0.10 & 0.03 & &-0.02 & 0.00 & 0.00 & -0.01 \\
   7500 & 4       & 0.06 & 0.03 & 0.08 & 0.08 & 0.04 & &-0.02 & 0.00 & 0.00 & -0.01 \\
   8000 & 4       & 0.05 & 0.01 & 0.06 & 0.07 & 0.05 & &-0.02 & 0.00 & 0.00 & -0.01 \\
   8500 & 4       & 0.06 & 0.00 & 0.05 & 0.07 & 0.08 & &-0.02 & 0.00 & 0.00 & -0.01 \\
   6500 & 3       & 0.01 & 0.10 & 0.14 & 0.12 & 0.01 & &-0.02 & -0.01 & -0.01 & -0.02 \\
   7000 & 3       & 0.03 & 0.08 & 0.12 & 0.11 & 0.02 & &-0.02 & -0.01 & 0.00 & -0.02 \\
   7500 & 3       & 0.06 & 0.05 & 0.10 & 0.11 & 0.06 & &-0.02 & 0.00 & 0.00 & -0.02 \\
   8000 & 3       & 0.10 & 0.06 & 0.11 & 0.14 & 0.14 & &-0.02 & 0.00 & 0.00 & -0.02 \\
   8500 & 3       & 0.13 & 0.06 & 0.12 & 0.18 & 0.20 & &-0.03 & 0.00 & 0.00 & -0.02 \\
\hline
\end{tabular}
\end{center}
\end{table}

\section{Departures from LTE depending on stellar parameters}

The non-LTE calculations were performed for the small grid of model atmospheres with stellar parameters characteristic of early F to late A-type stars: $\Teff  = 6500-8500$~K, $\logg = 3$ and 4. In this stellar parameter range, the general behavior of the departure coefficients of the Fe~I and Fe~II levels is independent of effective temperature and surface gravity and very similar to that shown in Fig.\,\ref{bf_8000}. Non-LTE leads to weakening the Fe~I lines and to the opposite effect for Fe~II. Table\,\ref{tab_corr} presents the non-LTE abundance corrections for the representative lines of Fe~I and Fe~II with various \Eexc. 

The non-LTE effects are only minor for Fe~II, such that the abundance correction does not exceed 0.03~dex in absolute value. For Fe~I in the $\logg = 4$ models, $\Delta{\rm NLTE} \le 0.12$~dex and, on average, decreases towards higher $\Teff$. For the lower gravity models, $\Delta{\rm NLTE} \le 0.20$~dex and, on average, grows towards higher $\Teff$. It is worth noting that the non-LTE abundance correction is substantially different for different lines, at any given temperature. 

Our theoretical results were compared with the non-LTE predictions of Rentzsch-Holm (1996) in common stellar parameter range, $\Teff$ = 7000 -- 8500~K, $\logg = 4$ and 3.5. 
There is no discrepancy for the Fe~II lines, while the departures from LTE for Fe~I are stronger in Rentzsch-Holm (1996) than in this study. 
For example, in the model 8500/4, Rentzsch-Holm (1996) gave the mean non-LTE correction of 0.17~dex which is 0.1~dex larger than our value. In contrast to our results, the non-LTE correction from Rentzsch-Holm (1996) grows towards higher $\Teff$ independent of stellar surface gravity. In both studies, the departures from LTE grow towards lower gravity, however, with $\logg = 3$, we obtain the smaller non-LTE corrections than Rentzsch-Holm (1996) did with $\logg = 3.5$. The difference between two studies is in the use of different model atoms. Our model atom of Fe~I is much more complete than that of Rentzsch-Holm (1996), where only 79 terms of Fe~I are included and most high-excitation levels with \Eexc\,$> 6$~eV are absent. This explains why the overionization of Fe~I is stronger in the Rentzsch-Holm (1996) calculations. 
 
\section{Conclusions}

\begin{itemize}
\item Completeness of the model atom is important for
correct calculation of the statistical equilibrium of iron.
\item In the stellar parameter range characteristic of middle F to middle A-type stars, LTE underestimates the element abundance derived from Fe~I lines, by 0.02 -- 0.1~dex for dwarfs depending on the line and stellar temperature and by up to 0.20~dex for giants.
\item LTE is as good as non-LTE for Fe~II. 
\item The uncertainty in available $gf-$values for visible lines of Fe~I and Fe~II is uncomfortably too high, such that solar and stellar iron abundances cannot be determined with the accuracy better than 0.1~dex. The astrophysical community shall ask atomic spectroscopists for new accurate measurements. 
\end{itemize}

\begin{acknowledgements}
This research was supported by the Russian Foundation for Basic Research (08-02-00469-a), the Russian Federal Agency on Science and Innovation (02.740.11.0247), and the Presidium RAS Programme ``Origin, structure, and evolution of the Universe objects'' (P19). 
\end{acknowledgements}

\end{document}